\documentclass[showpacs,showkeys,twocolumn,preprintnumbers,nofootinbib,groupedaddress,superscriptaddress,amsmath,amssymb]{revtex4-1}

\usepackage{graphicx}
\usepackage{dcolumn} 
\usepackage{bm}
\usepackage{amssymb}
\usepackage{amsmath}
\usepackage{epsfig}    
\usepackage{color}
\usepackage{slashed}

\begin{document} 

\title{Radiative Seesaw Mechanism for Charged Leptons}
\preprint{OU-HET-1096}

\author{Cheng-Wei Chiang}
\email{chengwei@phys.ntu.edu.tw}
\affiliation{Department of Physics, National Taiwan University, Taipei, Taiwan 10617, R.O.C.}
\affiliation{Physics Division, National Center for Theoretical Sciences, Taipei, Taiwan 10617, R.O.C.}

\author{Kei Yagyu}
\email{yagyu@het.phys.sci.osaka-u.ac.jp}
\affiliation{Department of Physics, Osaka University, Toyonaka, Osaka 560-0043, Japan}

\begin{abstract}

We discuss a mechanism where charged lepton masses are derived from one-loop diagrams mediated by particles in a dark sector including a dark matter candidate. 
We focus on a scenario where the muon and electron masses are generated at one loop with new ${\cal O}(1)$ Yukawa couplings. 
The measured muon anomalous magnetic dipole moment, $(g-2)_\mu$, can be explained in this framework. 
As an important prediction, the muon and electron Yukawa couplings can deviate significantly from their standard model predictions, and such deviations can be tested at High-Luminosity LHC and future $e^+e^-$ colliders. 

\end{abstract}
\maketitle

\noindent
{\it Introduction ---}
After the discovery of 125-GeV Higgs boson, direct evidence for the existence of the 
bottom~\cite{Aaboud:2018zhk,Sirunyan:2018kst} and tau~\cite{Aad:2015vsa,Sirunyan:2017khh} Yukawa couplings has been found at the LHC. 
In addition, the top Yukawa coupling has been indirectly probed from the gluon fusion production of Higgs boson and its diphoton decay~\cite{Aaboud:2018xdt,Mondal:2018hxy}. 
So far, these Yukawa couplings measured at the LHC are consistent with the Standard Model (SM) predictions within the errors, typically of a few $\times 10\%$ at $1\sigma$ level~\cite{Aad:2019mbh,Sirunyan:2018koj}. 
These facts suggest that the origin of mass in the third generation fermions can indeed be successfully described by the Yukawa interactions in the SM.  
Besides, the Higgs to dimuon decay has recently been observed at 2$\sigma$ and 3$\sigma$ levels at the ATLAS~\cite{Aad:2020xfq} and CMS~\cite{Sirunyan:2020two} experiments, respectively. 
Therefore, it is quite timely to scrutinize the origin of mass for first and second generation fermions.

In the SM, the large hierarchy in the assumed Yukawa couplings of charged fermions causes the flavor problem.  For example, the Yukawa couplings of the top quark and the electron differ by about five orders of magnitude. 
It is thus reasonable to suspect that some mechanism other than the usual Higgs-Yukawa interaction is at work to naturally explain the mass of light fermions.

One interesting idea is that light charged fermion masses are radiatively induced, and various models along this line had been proposed decades ago 
(see, for example, the review article~\cite{Babu:1989fg} and recent models after the Higgs discovery~\cite{Ma:2013mga,Gabrielli:2016vbb,Ma:2020wjc,Baker:2021yli}). 
More recently, Ma had proposed a scenario where particles in a dark sector, including a dark matter (DM) candidate, were introduced such that the mass generation is realized at loop level, with a concrete model constructed in Ref.~\cite{Ma:2013mga}. 
In this scenario, the Yukawa couplings for light fermions can be schematically expressed at $N$-loop level as 
\begin{align}
y_f = \left(\frac{1}{16\pi^2}\right)^N\times \frac{v}{M} \times Y_{\rm new} + {\cal O}(M^{-2}), 
\end{align}
where $v(\simeq 246$~GeV) is the vacuum expectation value (VEV) of the Higgs field $H$ ($\langle H \rangle = v/\sqrt{2}$), 
$M$ is the mass of the heaviest particle in the dark sector, and $Y_{\rm new}$ denotes a product of new couplings. 
Therefore, small fermion masses can naturally be explained by ${\cal O}(1)$ coupling $Y_{\rm new}$ with properly chosen $M$ and/or $N$.

\begin{figure}[t]
\begin{center}
 \includegraphics[width=70mm]{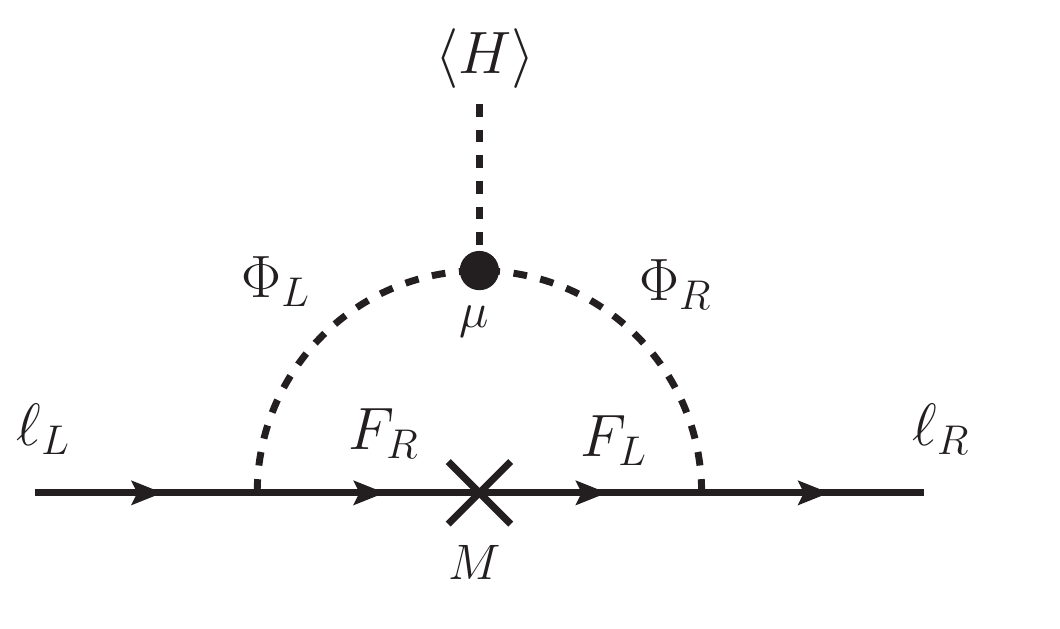}
   \caption{Radiative seesaw generation of charged lepton mass. }
   \label{fig:diagram}
\end{center}
\end{figure}

In this Letter, we further explore the idea of radiative seesaw mechanism for charged fermions and, in particular, focus on the scenario where the masses of muon and electron are generated at one-loop level, with particles in the dark sector running in the loop, as shown in Fig.~\ref{fig:diagram}.  Moreover, we study how the anomalous magnetic dipole moments, $(g-2)_\ell$, and the effective Yukawa couplings of these leptons are modified and can be tested in the near future.

\noindent
{\it Model ---}
We consider a model where the SM sector is distinguished from a dark sector by their respectively even and odd charges under an exact $Z_2$ symmetry.  The dark sector is composed of vector-like fermions $F^\ell$ and a pair of scalar fields ($\Phi_{L}$, $\Phi_{R}$), and the lightest $Z_2$-odd particle serves as a DM candidate.
In order to forbid tree-level mass for muon/electron, we further introduce a $Z_2'$ symmetry
under which the left- (right-) handed muon/electron are assigned as $Z_2$-even (odd). 
Finally, we impose a global $U(1)_\ell$ symmetry to avoid lepton flavor-violating (LFV) processes.

\begin{table}[t]
\begin{center}
\begin{tabular}{|c||ccc|ccc|}\hline
          & \multicolumn{3}{c|}{Fermions}  &  \multicolumn{3}{c|}{Scalars} \\\hline\hline
 Fields   &  $(L_L^e,L_L^\mu)$       & $(e_R,\mu_R)$     & $(F_{L/R}^e,F_{L/R}^\mu)$   & $H$     & $\Phi_{L}$ & $\Phi_{R}$  \\\hline\hline
$SU(2)_I$ &  ${\bm 2}$   & ${\bm 1}$ & $I_F$ & ${\bm 2}$ & $I_{L}$ & $I_{R}$   \\\hline
$U(1)_Y$  &  $-1/2$      & $-1$      & $Y_F$ & 1/2 & $Y_{L}$ & $Y_{R}$  \\\hline
$U(1)_\ell$  & $(q_e,q_\mu)$ &  $(q_e,q_\mu)$  & $(q_e,q_\mu)$ & $0$ & $0$& $0$ \\\hline
$Z_2$  & $(+,+)$ &  $(+,+)$ & $(-,-)$  & $+$ & $-$& $-$  \\\hline
$Z_2'$ & $(+,+)$ &  $(-,-)$ & $(+/-,+/-)$  & $+$ & $-$& $-$\\\hline
\end{tabular}
\caption{Charges of various particles under the $SU(2)_I\times U(1)_Y \times U(1)_\ell\times Z_2 \times Z_2'$ symmetry.  The unlisted $\tau$ lepton and quarks are not charged under $U(1)_\ell$ and are even under both $Z_2 \times Z_2'$. }
\label{particle}
\end{center}
\end{table}

Table~\ref{particle} summarizes the relevant particle content to realize our scenario mentioned above. 
In the fermion sector, two vector-like fermions $F^e$ and $F^\mu$ are separately introduced for the one-loop electron and muon masses without inducing dangerous LFV processes. 
In this table, $I_{F,L,R}$ and $Y_{F,L,R}$ are respectively the weak isospins and the hypercharges for the new particles.  To construct the required Yukawa interactions [Eq.~(\ref{eq:yukawa}) below], $Y_{L,R}$ are uniquely determined for a given $Y_F$ as: 
\begin{align}
Y_L = - 1/2 - Y_F,\quad Y_{R} = - 1 - Y_F. 
\label{eq:yl}
\end{align}
Possible combinations of the $SU(2)_I\times U(1)_Y$ charges for particles in the dark sector are listed in Table~\ref{particle2}. 
We note that models without neutral components in the dark sector [e.g., $(I_F,Y_F)=({\bm 1},1)$] are excluded because these models provide stable charged particles.

\begin{table}[t]
\begin{center}
\begin{tabular}{|c||ccc|}\hline
 $(I_F,Y_F)$     & $(I_{L},Y_{L})$ &  $(I_{R},Y_{R})$  & Sign of $\Delta a_\ell$  \\\hline\hline
$({\bm 1}, 0)$  & $({\bm 2}, -1/2)$ &  $({\bm 1}, -1)$ &  $+$ \\\hline
$({\bm 1}, -1)$  & $({\bm 2}, 1/2)$ &  $({\bm 1}, 0)$ & $-$ \\\hline
$({\bm 2}, 1/2)$  & $({\bm 1}~\text{or}~{\bm 3}, -1)$ &  $({\bm 2}, -3/2)$ & $+$ \\\hline
$({\bm 2}, -1/2)$  & $({\bm 1}~\text{or}~{\bm 3}, 0)$ &  $({\bm 2}, -1/2)$ & $-$ or $\pm$
\\\hline
$({\bm 2}, -3/2)$  & $({\bm 1}~\text{or}~{\bm 3}, 1)$ &  $({\bm 2},  1/2)$ &$-$\\\hline
$({\bm 3}, 1)$  & $({\bm 2}, -3/2)$ &  $({\bm 3}, -2)$   &$+$ \\\hline
$({\bm 3}, 0)$  & $({\bm 2}, -1/2)$ &  $({\bm 3}, -1)$   & $\pm$\\\hline
$({\bm 3}, -1)$  & $({\bm 2}, 1/2)$ &  $({\bm 3}, 0)$   &$-$ \\\hline
$({\bm 3}, -2)$  & $({\bm 2}, 3/2)$ &  $({\bm 3}, 1)$   &$-$ \\\hline
\end{tabular}
\caption{Possible $SU(2)_I\times U(1)_Y$ charges for the particles in the dark sector up to $SU(2)_I$ triplets.
The last column lists the sign of $\Delta a_\ell$ coming from new physics contributions. }
\label{particle2}
\end{center}
\end{table}

The relevant terms in the Lagrangian for our discussions are given by 
\begin{align}
\begin{split}
{\cal L} \supset&
- y_\tau\overline{L_L^\tau} H\tau_R -\mu H\cdot \Phi_{L}^*\cdot \Phi_{R} 
- \sum_{\ell=e,\mu} \Big(M_\ell \overline{F_L^\ell} F_R^\ell 
\\
& +f_L^\ell \overline{L_L^\ell}\cdot\Phi_L \cdot F_R^\ell 
+ f_R^\ell \overline{\ell_R}\Phi_R \cdot F_L^\ell \Big)    + \text{h.c.}, 
\end{split}
\label{eq:yukawa}
\end{align}
where $\cdot$ denotes an appropriate $SU(2)_I$ contraction, and all the above couplings can be taken to be real without loss of generality. 
The masses of the vector-like fermions $M_\ell$ softly break the $Z_2'$ symmetry.  
The $\mu$ term plays an important role in generating the charged lepton mass.  After the electroweak symmetry breaking, i.e., $H$ develops a non-zero VEV, 
it gives rise to a mixing between $\Phi_{L}$ and $\Phi_{R}$, and connects the left-handed and the right-handed fields of muon/electron, as seen in Fig.~\ref{fig:diagram}. 
Thanks to the $U(1)_\ell$ symmetry, the summation over the index $\ell$ is flavor-diagonal.

The Feynman diagrams for new contributions to the deviation in $(g-2)_\ell$, defined as $\Delta a_\ell \equiv  a_{\ell}^{\rm Exp} - a_{\ell}^{\rm SM}$ for $\ell = e$, $\mu$, are given by attaching the external photon line to the internal charged scalar or fermion line in Fig.~\ref{fig:diagram}. 
Hence, both mass and $g-2$ for the muon/electron are proportional to $\mu f_L^\ell f_R^\ell$. 
In particular, the sign of $\Delta a_\ell$, as given in the last column of Table~\ref{particle2}, is correlated with the canonical sign of the charged lepton mass.

In the following, we focus on the simplest model: $(I_F,Y_F)\sim ({\bm 1}, 0)$,  $(I_{L},Y_{L})\sim ({\bm 2}, -1/2)$ and $(I_{R},Y_{R})\sim ({\bm 1}, -1)$, 
to illustrate our radiative seesaw mechanism for light charged fermion mass and its implications (see also Ref.~\cite{Baker:2021yli} for a similar consideration).  We write the component scalar fields as $\Phi_L = (\phi_L^0,\phi_L^-)^T$ and $\Phi_R = \phi_R^-$. 
The $\mu$ term induces a mixing between $\phi_L^\pm$ and $\phi_{R}^\pm$. 
Define the mass eigenstates of charged scalar fields through $(\phi_L^\pm , \phi_R^\pm)^T = R(\theta)(\phi_1^\pm , \phi_2^\pm)^T$, where $R$ is a $2\times 2$ orthogonal matrix and $\theta$ is the mixing angle. 
The particle content of this model is the same as the model with $Y_D = 0$ proposed in Ref.~\cite{Chen:2020tfr},
and the DM phenomenology in our model is the same.  
As already shown in Ref.~\cite{Chen:2020tfr}, if one chooses the scalar field $\text{Re}[\phi_L^0]$ as the DM candidate, there are solutions satisfying the currently observed relic abundance, $\Omega_{\rm DM}h^2 \simeq 0.12$~\cite{Aghanim:2018eyx} at around $m_{\rm DM} = 63$ GeV and $m_{\rm DM} = 80$~GeV. 
In addition, it has been shown that the magnitude of the DM-DM-Higgs coupling, $\lambda_{\text{DM}}\equiv (m_{\phi_I}^2 - m_{\rm DM}^2)/(2v^2) - \lambda_{HL}/2$ 
with $m_{\phi_I}$ being the mass of ${\rm Im} (\phi_L^0)$ and ${\cal L} \supset \lambda_{HL}|H|^2|\Phi_L|^2$, 
has to be of ${\cal O}(10^{-3})$ or smaller in order to avoid the most stringent constraint from direct detections, i.e., the XENON1T experiment~\cite{Aprile:2018dbl}. 
Under the above consideration, we take $m_{\rm DM}=63$~GeV and $\lambda_{\text{DM}}=10^{-3}$ as a benchmark point, and use it in the following discussions.

\noindent
{\it Radiative Seesaw Mass ---}
The muon and electron masses are calculated by
\begin{align}
m_\ell &= -\frac{f_L^\ell f_R^\ell \sin{2\theta}}{32\pi^2}M_\ell \left[F(x_1^2) - F(x_2^2)\right], \label{eq:ml_exact}  
\end{align}
where $F(x) = \frac{1}{2}\left(\frac{1 + x}{1 - x} - 1\right)\ln x$, and $x_i\equiv m_{\phi_i^\pm}/M_\ell$ ($i=1,2$).  Therefore, Eq.~\eqref{eq:ml_exact} would vanish if $\phi_{1,2}^\pm$ are degenerate in mass.
For $x_i \ll 1$, we obtain $m_\ell \simeq f_L^\ell f_R^\ell \sin{2\theta} \, (x_1^2\ln x_1^2 - x_2^2\ln x_2^2)/(32\pi^2)$. 
Consequently, the small charged lepton masses can be naturally explained by having $m_{\phi_i^\pm} = {\cal O}(100)$~GeV and $f_L^\ell f_R^\ell \sin{2\theta} ={\cal O}(1)$, with $M_\mu$ and $M_e$ being of order TeV and PeV, respectively. 
This result also indicates that the electron mass can be reproduced by having $M_{e}={\cal O}(1)$~TeV and a smaller coupling product $f_L^e f_R^e \sin{2\theta} = m_e/m_\mu \simeq 1/200$, in which case
our original motivation to naturally explain the tiny mass would be lost.

\noindent
{\it Anomalous Magnetic Dipole Moments ---}
Currently, the discrepancies between the SM predictions and the data of $(g-2)_\mu$~\cite{Abi2021} and $(g-2)_e$~\cite{Morel2020} are given by 
\begin{align}
\begin{split}
\Delta a_\mu &= (251 \pm 59)\times 10^{-11},
\\ 
\Delta a_e   &= (4.8 \pm 3.0)\times 10^{-13}, 
\end{split}
\label{eq:dae}
\end{align}
where the value of $\Delta a_\mu$ is given by the combined data from BNL E821 and the first result of FNAL.\footnote{
The latest lattice QCD calculation~\cite{Borsanyi:2020mff} for hadronic contributions to $(g-2)_\mu$ 
shows a result consistent with the experimental data, while it differs significantly from calculations based on the dispersion relation. }
The above value of $\Delta a_e$ is extracted with the latest determination of fine-structure constant $\alpha_{\rm em}$ by using rubidium atoms~\cite{Morel2020}. 
Another measurement of $\alpha_{\rm em}$ using cesium atoms two years earlier~\cite{Jegerlehner2018} had a significant discrepancy with most other data, and implied $\Delta a_e=(-8.7 \pm 3.6)\times 10^{-13}$.  We do not take this value into account in our analysis. 
In our model, the new contribution to $g-2$ is given by 
\begin{align}
\begin{split}
\Delta a_\ell &= \frac{2m_\ell^2}{M_{\ell}^2}\frac{G(x_1^2) - G(x_2^2) }{F(x_1^2) - F(x_2^2)}, 
\label{eq:da_app}
\end{split}
\end{align}
with $G(x) = \left( 1-x^2+2x \ln x \right) / \left[ 2(1-x)^3 \right]$. 
Note that the dependence of the coupling product $f_Lf_R\sin{2\theta}$ does not explicitly enter because it is implicitly included in $m_\ell$. 
In addition, the sign of $\Delta a_\ell$ is determined to be positive, as already given in Table~\ref{particle2}, because both $F$ and $G$ are monotonically decreasing functions.  Therefore, the size of $\Delta a_\ell$ is mainly determined by $m_\ell^2/M_{\ell}^2$.

We note in passing that the contributions of two-loop Barr-Zee diagrams, exchanging the 125-GeV Higgs boson $h$ and with $\phi_{1,2}^\pm$ running in the loop where the photon attaches, 
are typically two orders of magnitude smaller than the one given in Eq.~(\ref{eq:da_app}) with the typical parameter choice.  
Therefore, they are negligible in our discussions.

\begin{figure}[t]
\begin{center}
 \includegraphics[width=40mm]{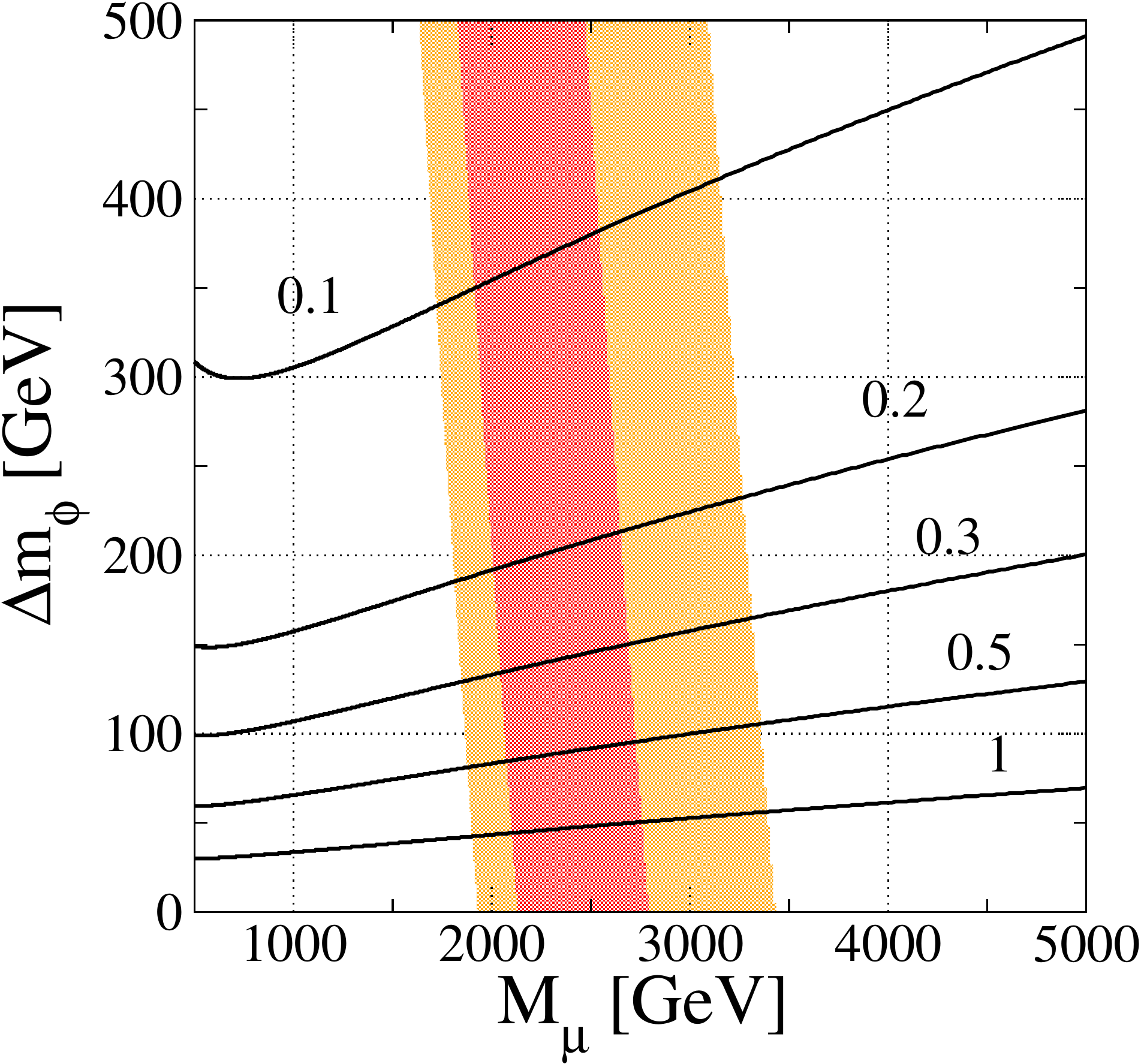}\hspace{3mm}
 \includegraphics[width=40mm]{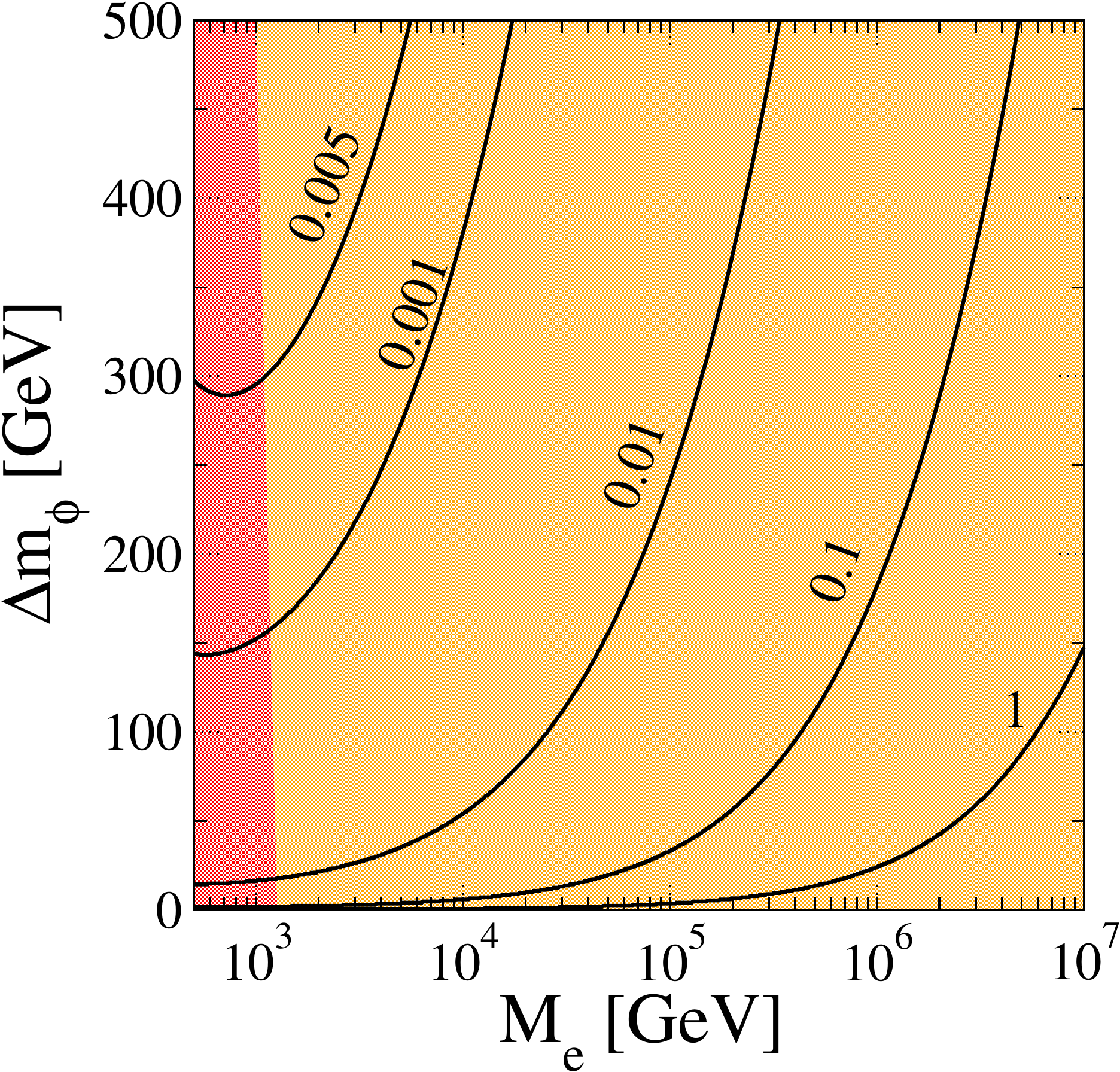}
\caption{Contour plots for the required values of $f_L^\ell f_R^\ell \sin{2\theta}$ to reproduce the correct charged lepton mass  in the $M_{\ell}$--$\Delta m_\phi$ 
plane, 
assuming $m_{\phi_1^\pm}=200$~GeV. The red (orange) region satisfies $\Delta a_\mu$ (left) and $\Delta a_e$ (right) at 1$\sigma$ (2$\sigma$) level. 
}
   \label{fig:g2}
\end{center}
\end{figure}

In Fig.~\ref{fig:g2}, we show the regions that can explain $(g-2)_\mu$ (left) and $(g-2)_e$ (right)
at $1\sigma$ (red) and $2\sigma$ (orange) levels in the $M_{\ell}$--$\Delta m_\phi$ plane for $m_{\phi_1^\pm}=200$~GeV with $\Delta m_\phi \equiv m_{\phi_2^\pm} - m_{\phi_1^\pm}$. 
For $(g-2)_e$, the $2\sigma$ region would cover negative values of $\Delta a_e$.  Therefore, the orange region only sets a lower limit on $M_{e}$, coming from the upper $2\sigma$ bound of $\Delta a_e$. 
The solid curves show the contours of the required value for the coupling product $f_L^\ell f_R^\ell\sin{2\theta}$ to reproduce the correct charged lepton mass. 
We see that $(g-2)_\mu$ can be explained within $1\sigma$ by taking $M_{\mu} = 2-3$~TeV and that the dependence on $\Delta m_\phi$ is mild.  
Similarly, $(g-2)_e$ can also be explained within $1\sigma$ by taking $M_{e}$ to be about 1~TeV. 
This can be simply understood by observing that $\Delta a_e/\Delta a_\mu \sim {\cal O}((M_\mu m_e)^2/(M_e m_\mu)^2)$. 
For $f_L^ef_R^e\sin{2\theta} = {\cal O}(1)$, the value of $M_{e}$ should be of order PeV as required by $m_e$.
In such a case, the new physics contribution to $\Delta a_e$ becomes negligibly small.

\noindent
{\it One-loop Yukawa Couplings ---}
Finally, we discuss the one-loop induced muon and electron Yukawa couplings which exhibit a different parameter dependence from their masses because now the scalar quartic couplings enter. 
These are calculated as 
\begin{align}
y_\ell
=& -\frac{m_\ell}{F(x_1^2) - F(x_2^2)}\Bigg[\sum_{i=1,2}
\sigma_i\lambda_{\phi_i^+\phi_i^-h}C_0(\phi_i^\pm,F^\ell,\phi_i^\pm)  \notag\\
& +\frac{\lambda_{\phi_1^\pm\phi_2^\mp h}}{t_{2\theta}} C_0(\phi_1^\pm,F^\ell,\phi_2^\pm)\Bigg],\label{eq:yell}
\end{align}
where $\sigma_i = +1~ (-1)$ for $i = 1~ (2)$, and $C_0(X,Y,Z)$ is the shorthand notation of the Passarino-Veltman three-point function~\cite{Passarino:1978jh}, 
$C_0(m_\ell^2,m_\ell^2,m_h^2;m_X,m_Y,m_Z)$.  
We define $\lambda_{\phi_i^+\phi_j^-h}$ $(i,j=1,2)$ as the coefficients of the $\phi_i^\pm$--$\phi_j^\mp$--$h$ vertices in the Lagrangian. 
When $\theta = \pi/4$, $x_1 \ll 1$ and $\Delta m_\phi \to 0$, the expression is approximated as
\begin{align}
 y_\ell &\simeq  \frac{m_\ell }{v}\frac{1}{|1 + \ln x_1^2|}
\left[|2 + \ln x_1^2|+\frac{v^2\lambda_0}{m_{\phi_1^\pm}^2} \right],  \label{eq:yell_app}
\end{align}
where $\lambda_0 = m_{\rm DM}^2/v^2  + \lambda_{\rm DM} - \lambda_{HR}/2$ with ${\cal L} \supset -\lambda_{HR}|H|^2|\Phi_R|^2$. 
It is convenient to define the scale factor $\kappa_\ell \equiv y_\ell/y_\ell^{\rm SM}$ which is obtained by 
dividing Eqs.~(\ref{eq:yell}) and (\ref{eq:yell_app}) by the factor of $m_\ell/v$. 
It is seen that for $\lambda_{HR} \simeq 0$, we have $\lambda_0 \sim 0.07$ (as we set $\lambda_{\rm DM} = 10^{-3}$ and $m_{\rm DM} = 63$ GeV for the DM phenomenology), and 
the second term of Eq.~(\ref{eq:yell_app}) can be neglected.  
In this case with $M_{\ell} = 2$~TeV and $m_{\phi_1^\pm}= 200$~GeV, $\kappa_\ell$ is about 0.75, meaning that the effective muon Yukawa coupling is slightly smaller than its value in the SM. 

In order to study theoretically well-defined regions, we impose the perturbativity bound which demands that magnitudes of the dimensionless couplings in the model, i.e., the 
three gauge couplings, the Yukawa couplings (top Yukawa and $f^{\ell} = f_{L}^{\ell} = f_{R}^{\ell}$) and the scalar quartic couplings are less than $4\pi$ up to 10~TeV. 
The scale dependence of these couplings are evaluated by using renormalization group equations (RGEs) at one-loop level. 
Initial values of the scalar couplings at the $m_Z$ scale are determined by inputting $m_h$, $m_{\phi_{1,2}^\pm}$, $m_{\rm DM}$, $m_{\phi_I}$, $\theta$, $\lambda_{\rm DM}$ and $\lambda_{HR}$, 
where $m_{\phi_I}$ is fixed such that the new contribution to the electroweak $T$ parameter~\cite{Peskin:1990zt} vanishes. 
There are the other three couplings for $Z_2$-odd scalar quartic vertices, and we take them to be zero at the $m_Z^{}$ scale. 
We note that these three couplings are only relevant to the calculation of the perturbativity bound using RGEs, and 
if we take them to have non-zero values the bound tends to be more severe. 

\begin{figure}[t]
\begin{center}
 \includegraphics[width=40mm]{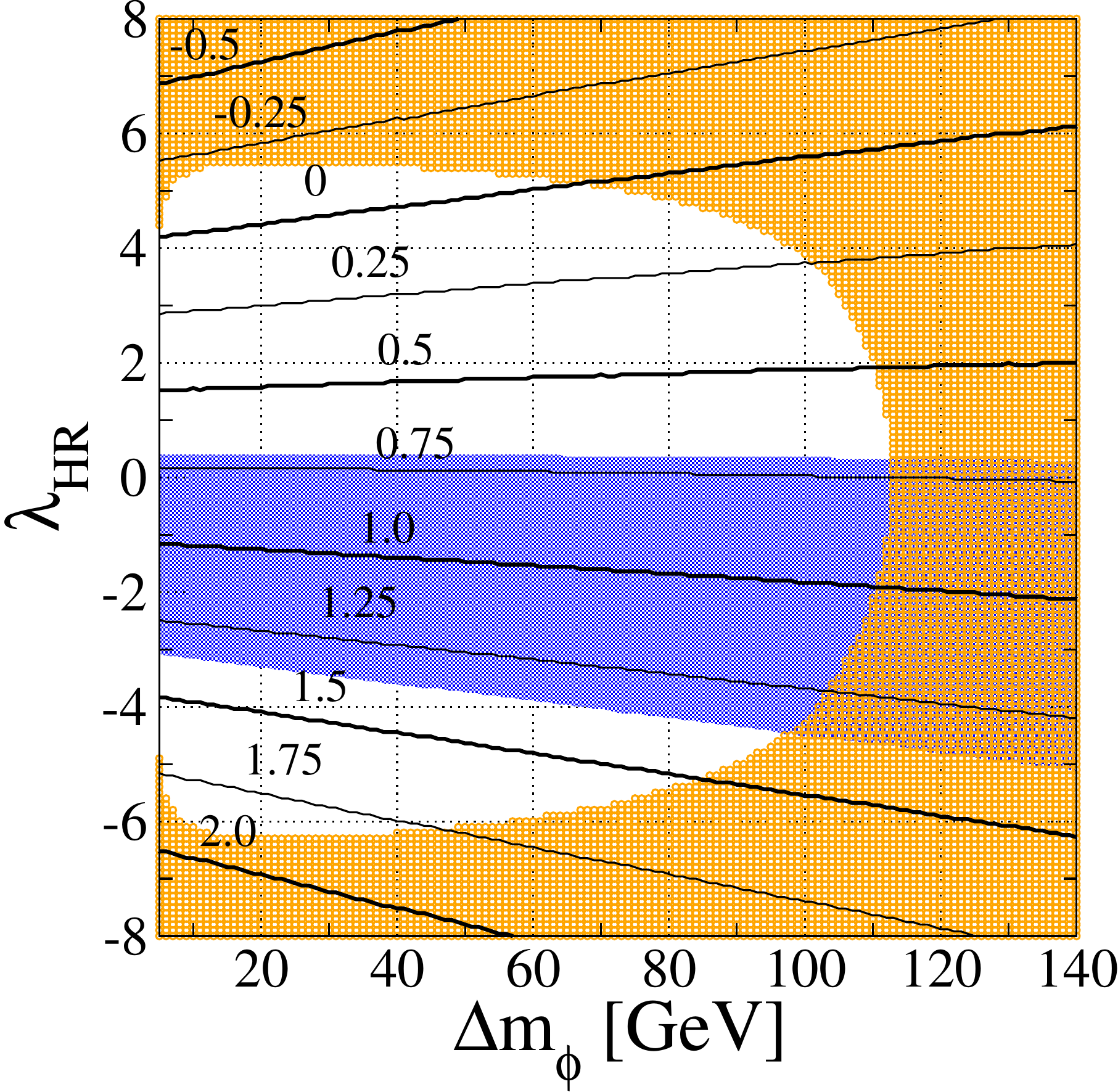}
 \includegraphics[width=40mm]{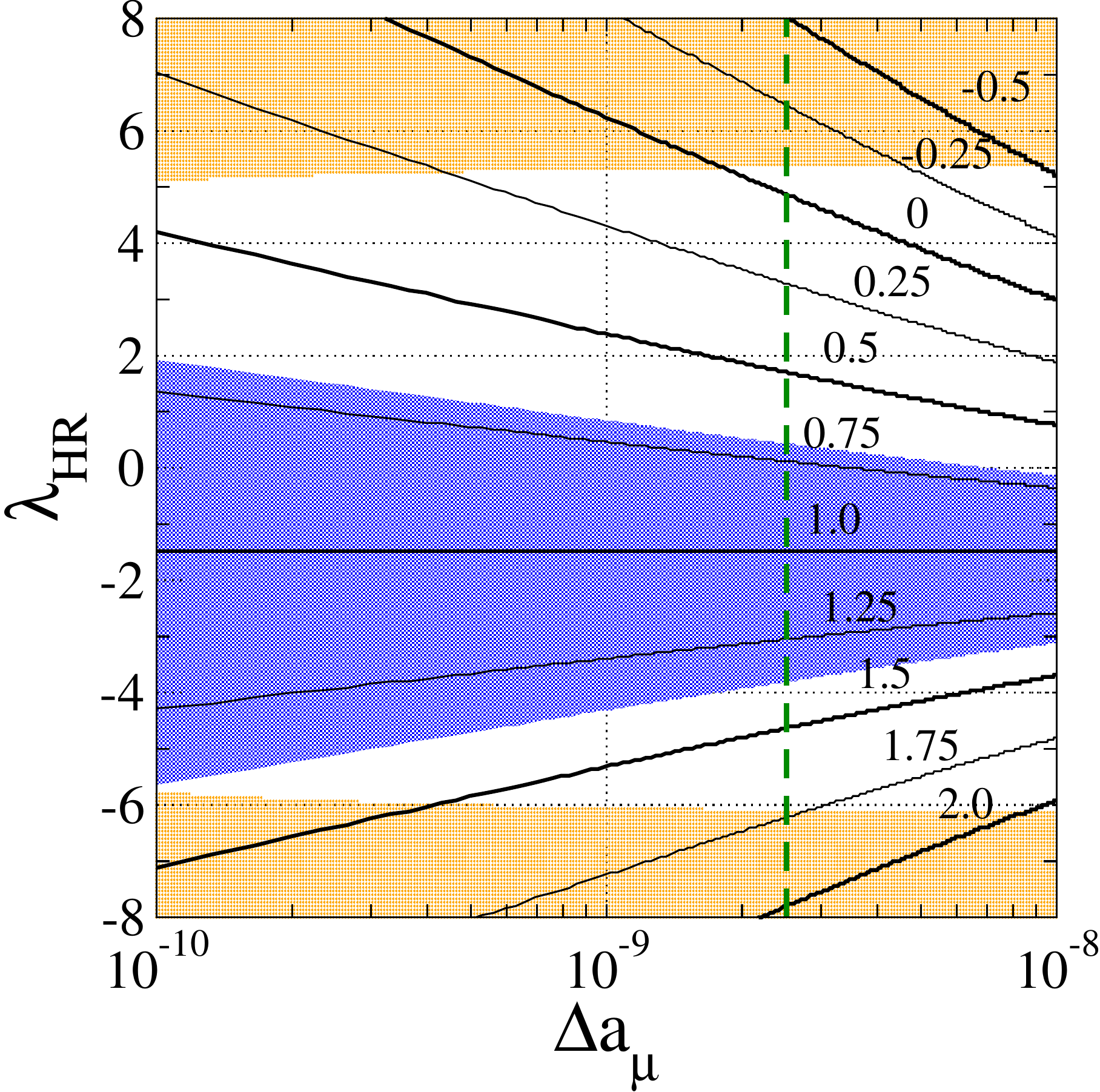}
   \caption{Contour plots for $\kappa_\mu$ on the $\Delta m_\phi$--$\lambda_{HR}$ (left) and $\Delta a_\mu$--$\lambda_{HR}$ (right) plane for $m_{\phi_1^\pm} = 200$~GeV, $\theta = \pi/4$ and $\Delta m_\phi = 50$ GeV (only right). 
The value of $M_{\mu}$ is fixed to reproduce the central value of $\Delta a_\mu$ given in Eq.~(\ref{eq:dae}) (left) and determined by the value of $x$-axis (right). 
The blue shaded region is allowed by the current measurement of the signal strength for $pp \to h \to \mu\mu$ at the LHC at $2\sigma$ level, while the orange region is excluded by the perturbativity bound.  
The vertical green dashed line in the right figure marks the current central value of $\Delta a_\mu$. }
   \label{fig:kappa}
\end{center}
\end{figure}

In Fig.~\ref{fig:kappa}, we show two contour plots of $\kappa_\mu$.
The value of $M_{\mu}$ is fixed to satisfy the central value of $\Delta a_\mu$ in the left plot, while it is determined by $\Delta a_\mu$ in the right plot.  
The blue shaded region is allowed by the current measurements of the signal strength $\mu_\mu$ for $pp \to h \to \mu\mu$ at the LHC at $2\sigma$ level, where the weighted average of ATLAS and CMS is $\mu_{\mu} = 1.19 \pm 0.35$. 
In our model, the production cross section does not change from the SM prediction and $h \to \mu\mu$ is a subdominant decay.  Therefore, the $\mu_\mu$ value is simply estimated as $\mu_\mu \simeq \kappa_\mu^2$. 
The orange region is excluded by the perturbativity bound. We check that bounds from perturbative unitarity and vacuum stability are much weaker than the perturbativity bound. 
As shown in the left plot, $\kappa_\mu$ monotonically decreases for larger $\lambda_{HR}$, while its dependence on $\Delta m_\phi$ is quite mild.

For future updates of $\Delta a_\mu$, we also show the dependence of $\kappa_\mu$ on $\Delta a_\mu$ in the right plot of Fig.~\ref{fig:kappa}.  We see that a larger deviation in $(g-2)_\mu$ tends to require a larger value of $|1 - \kappa_\mu|$ for a fixed value of $\lambda_{HR}$. 
We note that a similar deviation in the electron Yukawa coupling can be obtained when we take $M_e = {\cal O}(1)$~TeV.
From these results, we conclude that both the muon and electron Yukawa couplings can deviate significantly from their SM expectations by ${\cal O}(100\%)$ level without any contradiction with the theoretical bounds. 
Such a large deviation in the muon Yukawa coupling can be easily detected in future collider experiments. 
For example, at the High-Luminosity LHC and the International Linear Collider with 250~GeV and 2~ab$^{-1}$, $\kappa_\mu$ can be measured to the precision of about $7\%$~\cite{Cepeda:2019klc} and 5.6\%~\cite{Fujii:2017vwa} at $1\sigma$ level, respectively. 
We emphasize that our model generally predicts large deviations only in the muon and electron Yukawa couplings, while the other Higgs boson couplings, 
e.g., $\tau$ Yukawa coupling, quark Yukawa couplings and gauge couplings do not change from their SM values at tree level. 
Therefore, if a large deviation only in the muon Yukawa coupling is observed in future collider experiments, our model can offer a good explanation while 
providing a successful DM candidate, satisfying the required $\Delta a_\mu$ and naturally explaining the lepton masses.

{\it Discussions --} We briefly comment on the direct searches for scalar bosons in the dark sector. In our scenario, heavier scalar bosons
can decay into DM and a weak boson. Such a signature, i.e., mono-Z or mono-W is quite similar to that given in the inert doublet model, and 
the current constraint on the cross section is not so stringent (see e.g.,~\cite{Belanger:2015kga}). 
For the DM phenomenology, our DMs can mainly annihilate into a pair of muons and/or electrons, so that indirect DM searches from the measurements of $e^+e^-$ and/or $\mu^+\mu^-$ distributed in dwarf spheroidal galaxies at Fermi-LAT can be a useful probe. Current observations set a lower limit on the DM mass to be about 10 GeV~\cite{Ackermann:2015zua}. 
Therefore, our scenario is allowed by the above experiments.

To generate mass for active neutrinos, we need to extend the model by adding right-handed neutrinos, with which active neutrino masses can be generated at one loop using the so-called scotogenic mechanism proposed in Ref.~\cite{Ma:2006km}.  One potential problem is that we need to introduce an explicit breaking of the flavor $U(1)_\ell$ symmetry by Majorana masses of the right-handed neutrinos.  This can be avoided by introducing singlet scalar fields whose VEVs induce the breakdown of the $U(1)_\ell$ symmetry.  Detailed discussions on this issue are beyond the scope of the present Letter.


{\it Acknowledgments --} The authors would like to thank Takaaki Nomura, Osamu Seto and Koji Tsumura for giving useful comments on this project.  The works of CWC and KY were supported in part by Grant No.~MOST-108-2112-M-002-005-MY3 and the Grant-in-Aid for Early-Career Scientists, No.~19K14714, respectively.

\bibliography{references}

\end{document}